\documentclass[aps,prl,twocolumn,eqsecnum,amssymb,amsmath,showpacs,a4paper, superscriptaddress]{revtex4-1}
\usepackage{graphicx}
\usepackage{amsfonts}
\usepackage{amsmath}
\usepackage{hyperref}
\usepackage{units}
\usepackage[latin1]{inputenc}      		
\usepackage{color}
\usepackage{dcolumn}
\usepackage{bm}
\usepackage{float}
\usepackage{cleveref}
\crefname{section}{§}{§§}
\Crefname{section}{§}{§§}
\usepackage{mathtools}
\usepackage{chngcntr}
\counterwithout{equation}{section}

\renewcommand{\Im}{\textrm{Im}}

\newcommand{\om}{\omega}
\newcommand{\Om}{\Omega}

\renewcommand{\vec}[1]{\mathbf{#1}}
\newcommand{\be} {\begin{equation}}
\newcommand{\ee} {\end{equation}}
\newcommand{\bsub}{\begin{subequations}}
\newcommand{\esub}{\end{subequations}}
\newcommand{\bea}{\begin{eqnarray}}
\newcommand{\eea}{\end{eqnarray}}

\usepackage[normalem]{ulem}

\def\be{\begin{equation}}
\def\ee{\end{equation}}
\def\ba{\begin{align}}
\def\ea{\end{align}}

\begin{document}

\title{Black hole quasibound states from a draining bathtub vortex flow}

\author{Sam Patrick}
 \email{sampatrick31@googlemail.com}
 \affiliation{%
 School of Mathematical Sciences, University of Nottingham\\
 Nottingham, NG7 2FD, United Kingdom
}%
\author{Antonin Coutant}%
 \email{antonin.coutant@nottingham.ac.uk}
 \affiliation{%
 School of Mathematical Sciences, University of Nottingham\\
 Nottingham, NG7 2FD, United Kingdom
}%
\author{Maur\'icio Richartz}
\email{mauricio.richartz@ufabc.edu.br}
\affiliation{Centro de Matem\'atica, Computa\c{c}\~ao e Cogni\c{c}\~ao, Universidade Federal do ABC (UFABC), 09210-170 Santo Andr\'e, S\~ao Paulo, Brazil}
\author{Silke Weinfurtner}%
 \email{silkiest@gmail.com}
\affiliation{%
 School of Mathematical Sciences, University of Nottingham\\
 Nottingham, NG7 2FD, United Kingdom
}%

\date{\today}

\begin{abstract}
Quasinormal modes are a set of damped resonances that describe how an excited open system is driven back to equilibrium. In gravitational physics these modes characterise the ringdown of a perturbed black hole, e.g. following a binary black hole merger. A careful analysis of the ringdown spectrum reveals the properties of the black hole, such as its angular momentum and mass. In more complex gravitational systems the spectrum might depend on more parameters, and hence allows us to search for new physics. In this letter we present a hydrodynamic analogue of a rotating black hole, that illustrates how the presence of extra structure affects the quasinormal mode spectrum. The analogy is obtained by considering wave scattering on a draining bathtub vortex flow.
We show that due to vorticity of the background flow, the resulting field theory corresponds to a scalar field on an effective curved spacetime which acquires a local mass in the vortex core. The obtained quasinormal mode spectrum exhibits long-lived trapped modes, commonly known as quasibound states. Our findings can be tested in future experiments, building up on recent successful implementations of analogue rotating black holes.

\end{abstract}

\maketitle

\emph{Introduction.} The linear response of a perturbed black hole can be well-understood in terms of damped resonances called quasinormal modes (QNMs)~\cite{qnmreview,Konoplya:2011qq}. These damped modes possess a complex frequency whose real part corresponds to the oscillation frequency and whose imaginary part gives the lifetime. The QNM spectrum of a black hole is completely characterized by the black hole parameters, and does not depend on the initial conditions of the perturbations. With the on-going development of gravitational wave astronomy, significant efforts are being dedicated to use QNM spectra to reveal information about the structure of black holes, thereby allowing us to test General Relativity and alternative theories of Gravity~\cite{implications}. In this letter, we shall illustrate how the QNM spectrum can be drastically altered by additional structures in the context of Analogue Gravity. 

Analogue Gravity, pioneered by Unruh in 1981~\cite{unruh1981}, explores the possibility of testing gravitational effects in a broad variety of systems~\cite{AGreview}. For instance, it was shown in~\cite{surfacewave} that surface waves propagating on an inviscid, irrotational, and shallow fluid flow are equivalent to a scalar field propagating on an effective spacetime. This spacetime is completely determined by its propagation speed $c$ and the background fluid flow $\vec v$. In particular, it is possible to model a rotating black hole using an axisymetric fluid flow $\vec v(r) = v_r(r)\vec{e}_r + v_{\theta}(r)\vec{e}_{\theta}$. The fluid configuration will exhibit an ergosphere at $r=r_e$ if $|\vec v(r_e)|=c$ and an event horizon at $r=r_{\rm H}$ if $|v_r(r_{\rm H})|=c$. These two features are sufficient to give rise to many interesting effects that occur around astrophysical black holes, including Hawking radiation and superradiance~\cite{AGreview}. In particular, this allows us to experimentally test the universality and robustness of these effects. The past decade has seen an increase of interest in experimental realizations of analogue black holes, resulting in the measurement of both classical~\cite{vanc_expmnt,faccio_hwkrad,germain_watertank} and quantum~\cite{steinhauer} analogue Hawking radiation in (1+1)-dimensional systems. Experimental research on rotating, (2+1)-dimensional systems began more recently and is already bearing fruit. For example, superradiant amplification of surface waves was observed in a draining bathtub (DBT) vortex flow~\cite{SR_obsvn}. Rotating black hole analogues are also being explored using photon fluids~\cite{prain_exp}. 

In this work, we present a simplified model of a rotational DBT-type fluid flow motivated by realistic velocity profiles \cite{SR_obsvn,stepanyants,andersen} seen in experiments. Our model is also the analogue of a black hole with additional structure owing to the non-vanishing vorticity of the background flow in the centre of the vortex. Within our approximation, the vorticity causes perturbations to acquire a local effective mass close to the horizon. This is similar to certain studies of gravity, where a massless field can acquire a local mass by coupling to another field~\cite{Thomas1,Thomas2}. By studying the QNM spectrum, we find that our model also admits long-lived trapped modes in addition to usual QNMs. These are known in the literature as quasi-bound states (QBSs)~\cite{qnmreview} and are predicted to occur in particular for massive fields around Kerr black holes~\cite{QBS_kerr,dolandirac}.
Our findings are of relevance for both hydrodynamics and gravity. Our aim is to understand the consequences of additional structures on the QNM spectrum. The question arises in gravity when one introduces modifications to the usual Kerr metric~\cite{environment} as well as in fluid flows where the core structure of a vortex deviates from the irrotational case. 
With Analogue Gravity experiments gaining an increasing amount of momentum, it is likely that the challenge of measuring the QNM spectrum will be tackled in the near future. In order to perform such experiments, it is necessary to understand the dependence of the QNM spectrum on the specifics of the background flow.

\emph{Background flow and wave equation.}
The DBT vortex is a particularly simple hydrodynamic model based on the assumption of a highly symmetric flow. It is described by the irrotational and incompressible velocity profiles $v_{\theta}=C/r$ and $v_r = -D/r$ where $C$ (circulation) and $D$ (drain rate) are positive constants, and $(r,\theta)$ are the polar coordinates. The associated effective metric mimics a rotating black hole as it possesses a horizon at $r_{\rm H}=D/c$, and an ergosphere at $r_e = \sqrt{C^ 2 + D^ 2}/c$. Its QNM spectrum has been studied in the literature, see e.g.~\cite{QNM_DBT_stab,QNM_DBT,DBT_resonances,lepe2005}.
 
Even though the DBT model successfully describes the behaviour of the flow sufficiently far away from the centre, a vortex which forms under experimental conditions contains a core in which the flow is no longer irrotational. A more realistic formula for $v_{\theta}$ is given by the Rankine vortex (originally conceived for gases, this model is now widely used for all types of viscous fluids \cite{lautrup}),
\begin{equation} \label{rankine}
v_{\theta} = \frac{Cr}{r_0^2}\Theta(r_0-r) + \frac{C}{r}\Theta(r-r_0), 
\end{equation}
where $r_0$ is the radius of the vortex core and $\Theta$ is the Heaviside step function. An analytically amenable interpolation of this formula, which is a smooth function of $r$, was proposed by Rosenhead \cite{rosenhead} and later studied by Mih~\cite{mih2,vatistas,hydraulic},
\begin{equation}\label{mih}
v_{\theta} = \frac{Cr}{r_0^2+r^2}.
\end{equation}
Notice that there exists many more complicated models, offering a more accurate description of the vortex flow depending on the precise initial and boundary conditions of the flow~\cite{lautrup}. In this letter, we are interested in the main deviations introduced by a rotational core. Hence we will work with Eq.~\eqref{mih} as its analytic simplicity will lend itself to our frequency domain simulations. Since we are dealing with a two-dimensional axisymmetric model, the radial component is constrained by the incompressibility condition, which leads to the same radial velocity as in the DBT vortex, i.e.~$v_r=-D/r$. Using these velocity profiles, we investigate the QNM spectrum.

The equations governing an effective (2+1)-dimensional ideal fluid flow in the shallow water regime (single layer approximation~\cite{buhble}) are given by,
\begin{equation} \label{2D_Eqs}
\begin{split}
(\partial_t+\vec v\cdot\nabla)\vec v + g\nabla h & = 0, \\
(\partial_t+\vec v\cdot\nabla) h + h\nabla\cdot\vec v & = 0. 
\end{split}
\end{equation}
These equations are valid in the regime $h\ll L$, where $h$ is the height of the free surface and $L$ is the scale of variation in the $(r,\theta)$-plane. Perturbations $\textbf{u}$ to the background velocity can be expressed using a Helmholtz decomposition, $\textbf{u}=\nabla\phi+\tilde{\nabla}\psi$, where the cograd operator is defined as $\tilde{\nabla}=\vec{e}_z\wedge\nabla$ and $\vec e_z$ is the unit vector in the direction perpendicular to the (2+1) fluid ($\tilde{\nabla}$ can also be seen as the curl of the three dimensional vector field $\psi \vec e_z$). In the regime of short wavelengths $\lambda\ll L$ (but still shallow water $h \ll \lambda$), which amounts to a WKB approximation, the curl-free component obeys the wave equation, 
\begin{equation} \label{waveeqnmass}
(\partial_t+\vec v\cdot\nabla)^2\phi + \Omega_v^2\phi - c^2\nabla^2\phi = 0,
\end{equation}
and the other component is obtained through $(\partial_t+\vec v\cdot\nabla) \psi = - \Om_v \phi$. Moreover, the surface elevation $\delta h$ can be obtained from
\begin{equation} \label{eq:deltah}
g \delta h = - (\partial_t+\vec v\cdot\nabla) \phi + \Omega_v \psi  .
\end{equation}
In particular, the frequency content of $\delta h$ and $\phi$ will be identical. Eq.~\eqref{waveeqnmass} describes the propagation of a scalar field $\phi$ with a mass proportional to the background vorticity $\Omega_v=\tilde{\nabla}\cdot\vec v$. We note that wave equation~\eqref{waveeqnmass} becomes exact in two particular cases. The first is a solid body rotation $\vec v\propto r\vec{e}_{\theta}$ with $\Omega_v=\mathrm{const.}$, in which case the perturbations are called inertia gravity waves~\cite{buhble}. The second is an irrotational flow, in which case $\Omega_v = 0$ and the wave equation reduces to its standard form~\cite{gravwaves}. The problem of waves scattering on a Rankine-type vortex has been addressed in the literature, usually in the regime $|\vec v|\ll c$~\cite{kopiev,coste}. In our case, the effects we are interested in arise in the regime where $|\vec v|$ is of the order of $c$.

Since the velocity profile we assume is axisymmetric and stationary, Eq.~\eqref{waveeqnmass} can be solved by separation of variables. Hence, a generic perturbation can be written as a sum $\phi = \sum_{\om m} \phi_{\omega m}(r)\exp(im\theta-i\omega t)/\sqrt{r}$, where $m$ is the azimuthal number, and $\om$ the frequency. To simplify, one can perform a Boyer-Lindquist type transformation (see e.g.~\citep{DBT_resonances}) and introduce a radial tortoise coordinate $r_*$ through $dr_*=cdr/(c^2-v_r^2)$. Using this coordinate, the horizon is located at $r_* \to -\infty$, while at large $r$, we have $r_* \sim r$. The wave equation for a single frequency and azimuthal number reduces to,
\begin{equation} \label{waveeq2}
-\partial_{r_*}^2\phi_{\omega m} + V(r)\phi_{\omega m} = 0,
\end{equation}
\noindent where the effective potential $V$ is given by,
\begin{equation} \label{potential}
\begin{split} 
V(r) = -\left( \omega - \frac{mv_{\theta}}{r}\right)^2 + \left(c^2-\frac{D^2}{r^2}\right)\times & \\
\left(\frac{m^2-\frac{1}{4}}{r^2} + \frac{5D^2}{4c^2r^4}+\frac{\Omega_v^2}{c^2}\right) & .
\end{split}
\end{equation}
This differs from the usual potential since we are now using Eq.~\eqref{mih} for $v_\theta$ and the non-vanishing vorticity contributes the term $\Omega_v^2$, where $\Omega_v = \partial_r(rv_{\theta})/r$. Since the potential is symmetric under the transformation \{$\omega\rightarrow-\omega$, $m\rightarrow-m$\}, we restrict ourselves to $\mathrm{Re}\left(\omega\right)>0$ in the frequency domain. Hence, co- (counter-) rotating waves are defined by $m>0$ ($m<0$).

The QNM boundary conditions are that the wave is purely in-going on the horizon and purely out-going at spatial infinity. By solving  Eq.~\eqref{waveeq2} in the corresponding limits $r_* \rightarrow -\infty$ and $r_* \rightarrow \infty$, the boundary conditions can be expressed as,
\be \label{bcs}
\phi_{\omega m} \rightarrow \begin{cases}
 A_{\infty} e^{i \omega r_*}, \qquad &r_*\rightarrow \infty, \\
 A_{\mathrm{H}} e^{-i (\omega - m \Om_{\rm H}) r_*}, \qquad &r_*\rightarrow -\infty,
  \end{cases}
\ee 
where $A_{\infty}$ and $A_{\mathrm{H}}$ are constants, and $\Omega_{\mathrm{H}}$ is the angular frequency $\Omega_{\mathrm{H}}=(v_{\theta}/r)|_{r=r_{\mathrm{H}}}$ at the horizon. \\

\emph{Computing quasinormal modes.}
Solving the wave equation subject to the boundary conditions \eqref{bcs} selects a discrete set of complex frequencies $\om = \om^\mathrm{R} + i \Gamma$. Several methods are known in the literature to accomplish this objective~\cite{qnmreview}. We study the problem using three distinct methods. We start with a WKB method which allows us to make the distinction between the QNMs and QBSs and approximate their frequencies. Next, we implement a frequency domain simulation using a continued fraction method which allows us to accurately compute the quasinormal frequencies to 6-digit precision. Our last approach is a direct time domain simulation of Eq.~\eqref{waveeqnmass}, giving the time evolution of an initial perturbation of the vortex flow. Such a time evolution could be directly tested in future experiments. As we show, all three methods give consistent results. (In addition, by setting $r_0 = 0$, we have checked that all three methods reproduce the irrotational QNM spectrum for the DBT profile~\cite{Cardoso04,DBT_resonances}.)

\begin{figure} [!t]
\includegraphics[width=\linewidth]{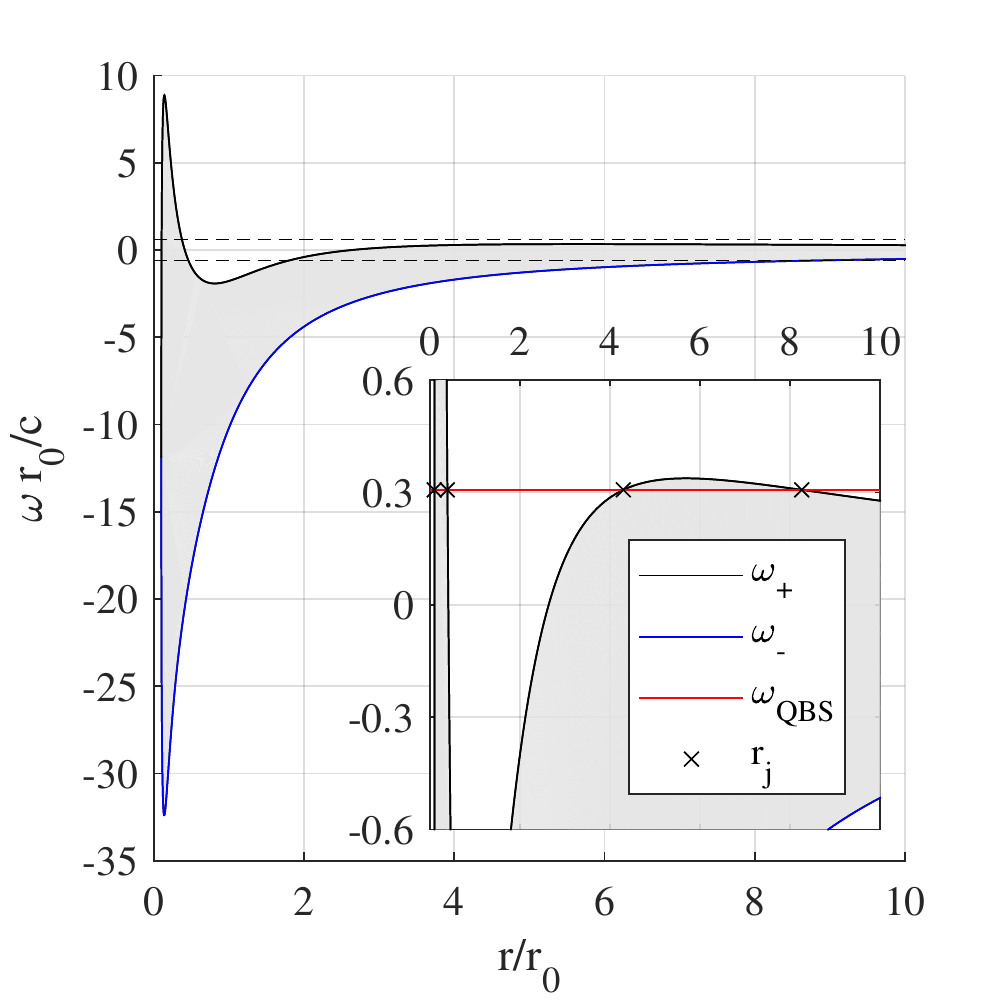}
\caption{We plot the functions $\omega_+(r)$ (black curve) and $\omega_-(r)$ (blue curve) defined in \eqref{ompm} for the dimensionless parameters $C/cr_0=3,D/cr_0=0.1$ and for $m=-4$ mode. The region between the dashed lines is shown in the embedded figure to highlight the presence of the smaller peak. The QNM frequencies are approximately those satisfying $\omega_{\pm}=\omega$, $\omega_{\pm}'=0$. A QBS frequency is shown for illustrative purposes. The turning points of the QBS are the $r_j$'s with $j=1...4$ from left to right in the figure.} \label{fig:potential}
\end{figure}

\emph{WKB method.} 
In the WKB regime (which is valid in particular for $|m|\gg1$) solutions are accurately described by a wave $\phi_{\omega m} \sim A \exp(i \int k(r_*) dr_*)$, where $A$ is a slowly varying amplitude and $k(r_*)$ is the local wavenumber which obeys the Hamilton-Jacobi equation, 
\be
k^2(r_*) = - V(r_*) \doteq (\om - \om_+(r_*)) (\om - \om_-(r_*)),
\ee 
with 
\be \label{ompm}
\om_\pm
=\frac{mv_{\theta}}{r} \pm \sqrt{\left(c^2-\frac{D^2}{r^2}\right) \left(\frac{m^2-\frac{1}{4}}{r^2} + \frac{5D^2}
{4c^2r^4}+\frac{\Omega_v^2}{c^2}\right)},
\ee
where $r$ is understood as a function of $r_*$.
The two functions $\om_\pm$ conveniently represent the potential \eqref{potential} for varying $\om$ (see Fig.~\ref{fig:potential}). Indeed, at the level of the WKB approximation, we see that if $\om$ lies outside the range $[\om_-, \om_+]$, then $k^2>0$. Hence the solution is oscillatory and the wave propagates. On the contrary, if $\om$ is inside this range, where $k^2<0$, the solution is evanescent. Moreover, points for which $\om = \om_\pm$ correspond to turning points. If in addition, the point is a local extremum of $\om_\pm$, it is an equilibrium point. 
Since we restrict ourselves to $\om >0$, it is sufficient to consider the extrema of $\om_+$. 

Near a local maximum, waves can hover around the vortex analogue of a light ring~\cite{DBT_resonances,Awesome_paper}. The real part of the QNM frequencies can be approximated through the conditions $\om=\om_+$ at the location such that $\om_+'=0$, where prime denotes the derivative with respect to $r_*$. In the eikonal limit ($|m|\gg1$), this condition is fulfilled at a single ($m$ independent) radius $r_{\rm lr}$, which is the light ring. Its orbital frequency $\om_{\rm lr} = \om_+|_{r_* = r_{\rm lr}}$ governs the QNM spectrum according to the formula~\cite{CardosoLR}, 
\begin{equation} \label{qnm_wkb}
\om_{\rm QNM}= m \om_{\rm lr} - i\left(n+\frac{1}{2} \right) \sqrt{\frac{-2V''}{(\partial_\om V)^2}}, 
\end{equation}
where $n\in\mathbb{Z}$ is called the overtone number and the term under the square root is evaluated at $r=r_{\rm lr}$, $\om=\om_{\rm lr}$. In the potential of Eq.~\eqref{waveeq2}, there are two maxima, and hence two associated families of QNMs. 

In addition to the usual QNMs, the existence of two peaks in the scattering potential means that quasibound states (QBSs) can also exist in the system. These QBSs can be understood as trapped modes in the potential well that must tunnel across the peaks to decay. Hence, the characteristic life-times of these modes are significantly larger than that of QNMs.  

To estimate the frequency of these QBSs, we perform a scattering amplitude calculation using WKB modes everywhere except at the turning points. We then construct a global solution using connection formulas at the turning points. When a QBS exists, the scattering potential contains four turning points $(r_{*j})_{j=1..4}$ (see Fig.~\ref{fig:potential}). We define $I_d, I_1$, and $I_2$ as the ranges spanned by the dip and the two peaks respectively; for example, $I_d = [r_{*2},r_{*3}]$. The real part $\om^\mathrm{R}_n$ of the QBS frequencies are given by the Bohr-Sommerfeld condition,
\begin{equation} \label{BScond}
S_d(\omega^\mathrm{R}_n) = \pi(n+1/2), 
\end{equation}
where $S_d(\omega^\mathrm{R}_n) = \int_{I_d} \sqrt{|V(\omega^\mathrm{R}_n,r)|}dr_*$ is the WKB action evaluated over the range $I_d$ and $n\in\mathbb{Z}$ indexes the different energy levels in the dip. In contrast to QNMs, this relation is satisfied only by a finite number of $\omega^\mathrm{R}_n$'s.

Once the real part is determined, the imaginary part $\Gamma_n$ of a long-lived mode with $\Gamma_n\ll \omega^\mathrm{R}_n$ is given by,
\begin{equation} \label{LTgamma}
\Gamma_n = -\frac{T_1^2 \pm T_2^2}{4\partial_{\omega}S_d|_{\omega=\omega^\mathrm{R}_n}},
\end{equation}
where the transmission coefficients across the inner ($1$) and outer ($2$) potential barriers are given by $T_{1,2} = \exp(-S_{1,2})$  
in the WKB regime and $S_{1,2}$ are the actions evaluated over the ranges $I_{1,2}$. In Eq.~\eqref{LTgamma} we take the $-$ sign if $\om^\mathrm{R} < m \Om_H$ corresponding to a superradiant amplification on the inner barrier, and the $+$ sign otherwise. By comparing Eqs.~\eqref{qnm_wkb} and \eqref{LTgamma}, we see that the life-time of QBSs are typically exponentially large compared to that of QNMs. A full derivation of Eqs.~(\ref{BScond}-\ref{LTgamma}) is outlined in the Supplemental Material~\footnote{See Supplemental Material, which includes Refs.~\cite{berrymount,Froman,CoutantCF,leaverkerr,leaverrn,nollert,Cardoso04,gautschi,Cook14,Konoplya:2011qq}, for a detailed description of the WKB method, the continued fraction method, and our numerical simulations, including a comparison between them.}. \\

\begin{figure} [!t]
\includegraphics[width=\linewidth]{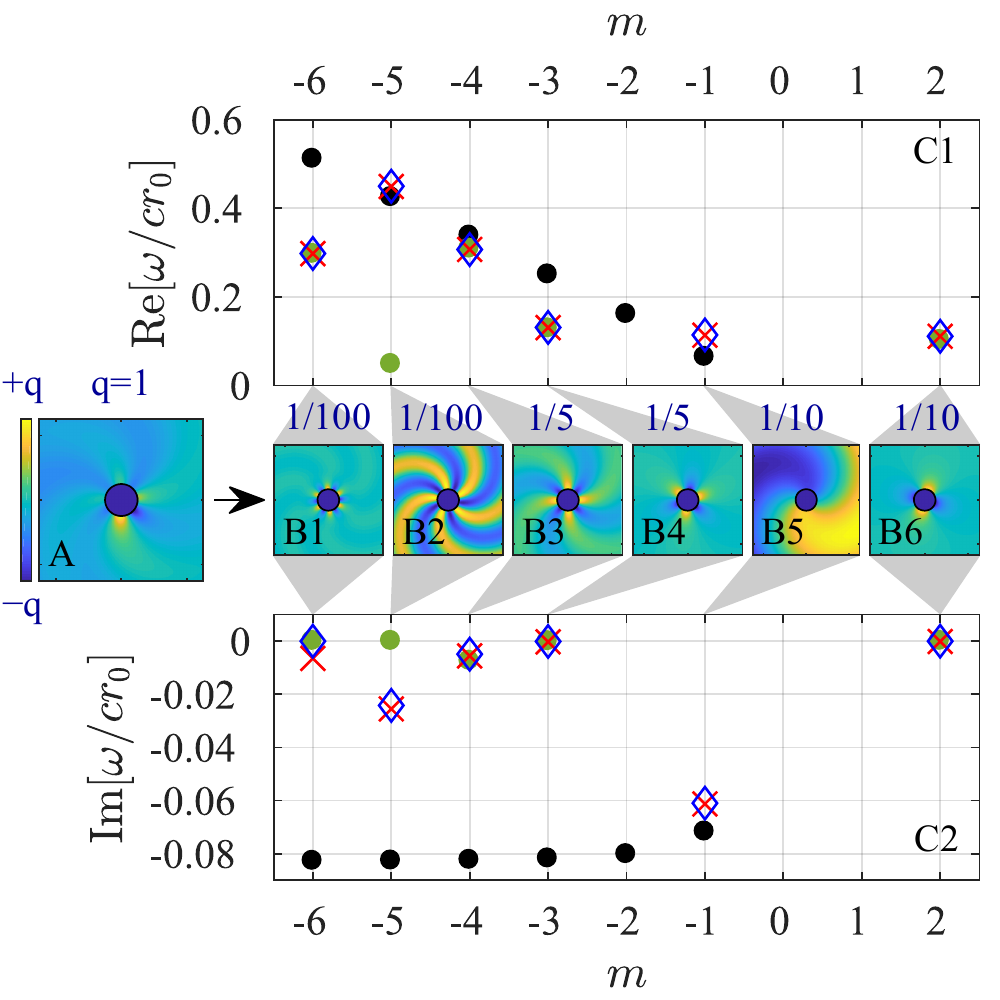}
\caption{
Results obtained for the fluid flow parameters $C/cr_0=3,D/cr_0=0.1$. In panel A we show the ringdown obtained from the MOL simulation of a initial gaussian wave packet after it has passed the vortex. In panels B1-B6 we show the decomposition of the ringdown signal onto an azimuthal basis, for $m\in \{-6,-5,-4,-3,-1,+2 \}$. We only display the values of $m$ for which the signal is above the noise level. To show the relative amplitudes of each component we refer to a common colour bar rescaled by a factor $q$ (indicated above each of the panels B1-B6), normalised to one for overall signal (panel A). 
In panels C1 and C2 we show the real and imaginary parts of the complex frequency spectrum obtained by the three methods: MOL (red crosses), CFM (blue diamonds) and WKB (green dots for QBSs and black dots for usual QNMs). The MOL and CFM are in excellent agreement. Moreover, by comparing the MOL/CFM with the WKB results we see that only $m=-1$ is a usual QNMs, while $m\in \{-6,-4,-3,+2 \}$ are QBSs. Note, that the distinction between usual QNMs and QBSs close to the peaks in the scattering potential is not sharply defined, as in the case of the $m=-5$ mode here.} \label{fig:results}
\end{figure}

\emph{Numerical simulations.} 
Our first numerical method is a direct time domain simulation of Eq.~\eqref{waveeqnmass} using a gaussian pulse parallel to  one of the cartesian axes as initial data. The wave equation is numerically integrated using the method of lines (MOL) and a 4th-order Runge-Kutta (RK4) algorithm. The quasinormal frequencies are extracted by performing a time Fourier transform of the signal once the initial pulse has passed. The second method is a frequency domain simulation implemented through a continued fraction method (CFM)~\cite{leaverkerr,nollert}. The two methods are detailed in the Supplemental Material~\cite{Note1}. \\

\emph{Results.} 
In Fig.~\ref{fig:results} we present our main results. We use the MOL simulations to reconstruct the full 2-dimensional pattern $\phi(r,\theta)$ (panel A) and its decomposition in azimuthal modes $\phi_m(r)e^{i m \theta}$ (panels B1-B5) at fixed time after the initial pulse has passed the vortex. We also present the real (panel C1) and imaginary (panel C2) components of complex frequency spectrum, obtained by the three different methods.  We observe an excellent agreement between the results from the MOL (red crosses) and the CFM (blue diamonds) simulations. The WKB approximation then allows us to identify the underlying structure of the spectrum, i.e.~whether the mode is a QNM (solid black dots) or a QBS (solid green dots). We show that most of the excited modes are QBSs, and as a result the ringdown has a significantly prolonged lifetime.

\emph{Conclusion.}
Our results show that the core structure of a vortex can significantly affect its QNM spectrum, providing a fluid analogue of the problem of spectral stability in black hole physics~\cite{environment}. Specifically, a decrease in the angular velocity in the vortex core creates a local minimum in the effective scattering potential which supports new resonances called quasibound states (the origin of these states is the same as for bound states of massive fields around Kerr black holes). Furthermore, we argued that the main effect of vorticity of the background on the propagation of waves can be encoded in a quantity which preserves the causal structure of the geometry, namely an effective local mass (see Eq.~\eqref{waveeqnmass}). This is similar to scalar-tensor theories of gravity, where scalar perturbations also possess a spatially varying effective mass. In addition, the conclusions drawn in this paper should be observable experimentally by analyzing the ringdown of a vortex flow. If the core of the vortex lies behind the horizon ($r_0 < r_{\rm H}$), then the observed spectrum will be close to that of an ideal DBT. On the contrary, if the core is large enough ($r_0 > r_{\rm H}$) our results show that the presence of QBSs will significantly alter the time response of the system by allowing long-lived modes to hover around the vortex. As a last remark, we notice that QBSs, and even bound states, will also appear if the analogue black hole is encompassed by a circular rigid boundary~\cite{Benone14}. In our letter, however, we have demonstrated that vorticity alone is sufficient to trap modes around the analogue black hole and produce QBSs. 

We end this letter by sketching a possible experimental setup to test our predictions. For this, we suggest a draining vortex flow similar to Ref.~[11] but shallower (such that the 2D effective description leading to \eqref{2D_Eqs} becomes accurate). To better control the rotation, a rotating plate can be added at the bottom of the tank with a drain hole in its center. The radius of the plate allows one to control $r_0$. Indeed, above it, viscous effects will impose a fluid velocity profile equal to that of the plate (that is, a solid body rotation as in the center of the Rankine vortex). By stimulating the vortex with a wave pulse, one will excite the QNMs and QBSs.

The ringdown we described above would manifest itself as rotating spirals imprinted on the free surface. In fact, such spirals are known to appear frequently around vortex flows (see Fig.~1 in Supplementary Material), for which a theoretical description seems to be lacking in the literature. We believe that the study of resonance frequencies could provide a fair description of this phenomenon.

\acknowledgements
\emph{Acknowledgements.}
The authors would like to thank Emanuele Berti, Thomas Sotiriou and Th\' eo Torres for illuminating discussions. This project has received funding from the European Union's Horizon 2020 research and innovation programme under the Marie Sk\l odowska-Curie grant agreement No 655524. M.R. acknowledges partial support from the S\~ao Paulo Research Foundation (FAPESP), Grant No. 2013/09357-9, and from the Fulbright Visiting Scholars Program. M.R. is also grateful to the University of Mississippi for hospitality while part of this research was being conducted. S.W. acknowledges financial support provided under the Royal Society University Research Fellow (UF120112), the Nottingham Advanced Research Fellow (A2RHS2), the Royal Society Project (RG130377) grants, the Royal Society Enhancement Grant (RGF/EA/180286) and the EPSRC Project Grant (EP/P00637X/1). SW acknowledges partial support from STFC consolidated grant No. ST/P000703/.

\bibliography{QNM-DBT.bbl}

\begin{thebibliography}{45}%
\makeatletter
\providecommand \@ifxundefined [1]{%
 \@ifx{#1\undefined}
}%
\providecommand \@ifnum [1]{%
 \ifnum #1\expandafter \@firstoftwo
 \else \expandafter \@secondoftwo
 \fi
}%
\providecommand \@ifx [1]{%
 \ifx #1\expandafter \@firstoftwo
 \else \expandafter \@secondoftwo
 \fi
}%
\providecommand \natexlab [1]{#1}%
\providecommand \enquote  [1]{``#1''}%
\providecommand \bibnamefont  [1]{#1}%
\providecommand \bibfnamefont [1]{#1}%
\providecommand \citenamefont [1]{#1}%
\providecommand \href@noop [0]{\@secondoftwo}%
\providecommand \href [0]{\begingroup \@sanitize@url \@href}%
\providecommand \@href[1]{\@@startlink{#1}\@@href}%
\providecommand \@@href[1]{\endgroup#1\@@endlink}%
\providecommand \@sanitize@url [0]{\catcode `\\12\catcode `\$12\catcode
  `\&12\catcode `\#12\catcode `\^12\catcode `\_12\catcode `\%12\relax}%
\providecommand \@@startlink[1]{}%
\providecommand \@@endlink[0]{}%
\providecommand \url  [0]{\begingroup\@sanitize@url \@url }%
\providecommand \@url [1]{\endgroup\@href {#1}{\urlprefix }}%
\providecommand \urlprefix  [0]{URL }%
\providecommand \Eprint [0]{\href }%
\providecommand \doibase [0]{http://dx.doi.org/}%
\providecommand \selectlanguage [0]{\@gobble}%
\providecommand \bibinfo  [0]{\@secondoftwo}%
\providecommand \bibfield  [0]{\@secondoftwo}%
\providecommand \translation [1]{[#1]}%
\providecommand \BibitemOpen [0]{}%
\providecommand \bibitemStop [0]{}%
\providecommand \bibitemNoStop [0]{.\EOS\space}%
\providecommand \EOS [0]{\spacefactor3000\relax}%
\providecommand \BibitemShut  [1]{\csname bibitem#1\endcsname}%
\let\auto@bib@innerbib\@empty
\bibitem [{\citenamefont {Berti}\ \emph {et~al.}(2009)\citenamefont {Berti},
  \citenamefont {Cardoso},\ and\ \citenamefont {Starinets}}]{qnmreview}%
  \BibitemOpen
  \bibfield  {author} {\bibinfo {author} {\bibfnamefont {E.}~\bibnamefont
  {Berti}}, \bibinfo {author} {\bibfnamefont {V.}~\bibnamefont {Cardoso}}, \
  and\ \bibinfo {author} {\bibfnamefont {A.~O.}\ \bibnamefont {Starinets}},\
  }\href@noop {} {\bibfield  {journal} {\bibinfo  {journal} {Classical and
  Quantum Gravity}\ }\textbf {\bibinfo {volume} {26}},\ \bibinfo {pages}
  {163001} (\bibinfo {year} {2009})}\BibitemShut {NoStop}%
\bibitem [{\citenamefont {Konoplya}\ and\ \citenamefont
  {Zhidenko}(2011)}]{Konoplya:2011qq}%
  \BibitemOpen
  \bibfield  {author} {\bibinfo {author} {\bibfnamefont {R.~A.}\ \bibnamefont
  {Konoplya}}\ and\ \bibinfo {author} {\bibfnamefont {A.}~\bibnamefont
  {Zhidenko}},\ }\href {\doibase 10.1103/RevModPhys.83.793} {\bibfield
  {journal} {\bibinfo  {journal} {Rev. Mod. Phys.}\ }\textbf {\bibinfo {volume}
  {83}},\ \bibinfo {pages} {793} (\bibinfo {year} {2011})},\ \Eprint
  {http://arxiv.org/abs/1102.4014} {arXiv:1102.4014 [gr-qc]} \BibitemShut
  {NoStop}%
\bibitem [{\citenamefont {Yunes}\ \emph {et~al.}(2016)\citenamefont {Yunes},
  \citenamefont {Yagi},\ and\ \citenamefont {Pretorius}}]{implications}%
  \BibitemOpen
  \bibfield  {author} {\bibinfo {author} {\bibfnamefont {N.}~\bibnamefont
  {Yunes}}, \bibinfo {author} {\bibfnamefont {K.}~\bibnamefont {Yagi}}, \ and\
  \bibinfo {author} {\bibfnamefont {F.}~\bibnamefont {Pretorius}},\ }\href@noop
  {} {\bibfield  {journal} {\bibinfo  {journal} {Physical review D}\ }\textbf
  {\bibinfo {volume} {94}},\ \bibinfo {pages} {084002} (\bibinfo {year}
  {2016})}\BibitemShut {NoStop}%
\bibitem [{\citenamefont {Unruh}(1981)}]{unruh1981}%
  \BibitemOpen
  \bibfield  {author} {\bibinfo {author} {\bibfnamefont {W.~G.}\ \bibnamefont
  {Unruh}},\ }\href@noop {} {\bibfield  {journal} {\bibinfo  {journal}
  {Physical Review Letters}\ }\textbf {\bibinfo {volume} {46}},\ \bibinfo
  {pages} {1351} (\bibinfo {year} {1981})}\BibitemShut {NoStop}%
\bibitem [{\citenamefont {Barcel{\'o}}\ \emph {et~al.}(2011)\citenamefont
  {Barcel{\'o}}, \citenamefont {Liberati},\ and\ \citenamefont
  {Visser}}]{AGreview}%
  \BibitemOpen
  \bibfield  {author} {\bibinfo {author} {\bibfnamefont {C.}~\bibnamefont
  {Barcel{\'o}}}, \bibinfo {author} {\bibfnamefont {S.}~\bibnamefont
  {Liberati}}, \ and\ \bibinfo {author} {\bibfnamefont {M.}~\bibnamefont
  {Visser}},\ }\href@noop {} {\bibfield  {journal} {\bibinfo  {journal} {Living
  reviews in relativity}\ }\textbf {\bibinfo {volume} {14}},\ \bibinfo {pages}
  {3} (\bibinfo {year} {2011})}\BibitemShut {NoStop}%
\bibitem [{\citenamefont {Sch{\"u}tzhold}\ and\ \citenamefont
  {Unruh}(2002{\natexlab{a}})}]{surfacewave}%
  \BibitemOpen
  \bibfield  {author} {\bibinfo {author} {\bibfnamefont {R.}~\bibnamefont
  {Sch{\"u}tzhold}}\ and\ \bibinfo {author} {\bibfnamefont {W.~G.}\
  \bibnamefont {Unruh}},\ }\href@noop {} {\bibfield  {journal} {\bibinfo
  {journal} {Physical Review D}\ }\textbf {\bibinfo {volume} {66}},\ \bibinfo
  {pages} {044019} (\bibinfo {year} {2002}{\natexlab{a}})}\BibitemShut
  {NoStop}%
\bibitem [{\citenamefont {Weinfurtner}\ \emph {et~al.}(2011)\citenamefont
  {Weinfurtner}, \citenamefont {Tedford}, \citenamefont {Penrice},
  \citenamefont {Unruh},\ and\ \citenamefont {Lawrence}}]{vanc_expmnt}%
  \BibitemOpen
  \bibfield  {author} {\bibinfo {author} {\bibfnamefont {S.}~\bibnamefont
  {Weinfurtner}}, \bibinfo {author} {\bibfnamefont {E.~W.}\ \bibnamefont
  {Tedford}}, \bibinfo {author} {\bibfnamefont {M.~C.}\ \bibnamefont
  {Penrice}}, \bibinfo {author} {\bibfnamefont {W.~G.}\ \bibnamefont {Unruh}},
  \ and\ \bibinfo {author} {\bibfnamefont {G.~A.}\ \bibnamefont {Lawrence}},\
  }\href@noop {} {\bibfield  {journal} {\bibinfo  {journal} {Physical review
  letters}\ }\textbf {\bibinfo {volume} {106}},\ \bibinfo {pages} {021302}
  (\bibinfo {year} {2011})}\BibitemShut {NoStop}%
\bibitem [{\citenamefont {Belgiorno}\ \emph {et~al.}(2010)\citenamefont
  {Belgiorno}, \citenamefont {Cacciatori}, \citenamefont {Clerici},
  \citenamefont {Gorini}, \citenamefont {Ortenzi}, \citenamefont {Rizzi},
  \citenamefont {Rubino}, \citenamefont {Sala},\ and\ \citenamefont
  {Faccio}}]{faccio_hwkrad}%
  \BibitemOpen
  \bibfield  {author} {\bibinfo {author} {\bibfnamefont {F.}~\bibnamefont
  {Belgiorno}}, \bibinfo {author} {\bibfnamefont {S.}~\bibnamefont
  {Cacciatori}}, \bibinfo {author} {\bibfnamefont {M.}~\bibnamefont {Clerici}},
  \bibinfo {author} {\bibfnamefont {V.}~\bibnamefont {Gorini}}, \bibinfo
  {author} {\bibfnamefont {G.}~\bibnamefont {Ortenzi}}, \bibinfo {author}
  {\bibfnamefont {L.}~\bibnamefont {Rizzi}}, \bibinfo {author} {\bibfnamefont
  {E.}~\bibnamefont {Rubino}}, \bibinfo {author} {\bibfnamefont
  {V.}~\bibnamefont {Sala}}, \ and\ \bibinfo {author} {\bibfnamefont
  {D.}~\bibnamefont {Faccio}},\ }\href@noop {} {\bibfield  {journal} {\bibinfo
  {journal} {Physical review letters}\ }\textbf {\bibinfo {volume} {105}},\
  \bibinfo {pages} {203901} (\bibinfo {year} {2010})}\BibitemShut {NoStop}%
\bibitem [{\citenamefont {Euv{\'e}}\ \emph {et~al.}(2016)\citenamefont
  {Euv{\'e}}, \citenamefont {Michel}, \citenamefont {Parentani}, \citenamefont
  {Philbin},\ and\ \citenamefont {Rousseaux}}]{germain_watertank}%
  \BibitemOpen
  \bibfield  {author} {\bibinfo {author} {\bibfnamefont {L.-P.}\ \bibnamefont
  {Euv{\'e}}}, \bibinfo {author} {\bibfnamefont {F.}~\bibnamefont {Michel}},
  \bibinfo {author} {\bibfnamefont {R.}~\bibnamefont {Parentani}}, \bibinfo
  {author} {\bibfnamefont {T.}~\bibnamefont {Philbin}}, \ and\ \bibinfo
  {author} {\bibfnamefont {G.}~\bibnamefont {Rousseaux}},\ }\href@noop {}
  {\bibfield  {journal} {\bibinfo  {journal} {Physical review letters}\
  }\textbf {\bibinfo {volume} {117}},\ \bibinfo {pages} {121301} (\bibinfo
  {year} {2016})}\BibitemShut {NoStop}%
\bibitem [{\citenamefont {Steinhauer}(2016)}]{steinhauer}%
  \BibitemOpen
  \bibfield  {author} {\bibinfo {author} {\bibfnamefont {J.}~\bibnamefont
  {Steinhauer}},\ }\href@noop {} {\bibfield  {journal} {\bibinfo  {journal}
  {Nature Physics}\ }\textbf {\bibinfo {volume} {12}},\ \bibinfo {pages} {959}
  (\bibinfo {year} {2016})}\BibitemShut {NoStop}%
\bibitem [{\citenamefont {Torres}\ \emph
  {et~al.}(2017{\natexlab{a}})\citenamefont {Torres}, \citenamefont {Patrick},
  \citenamefont {Coutant}, \citenamefont {Richartz}, \citenamefont {Tedford},\
  and\ \citenamefont {Weinfurtner}}]{SR_obsvn}%
  \BibitemOpen
  \bibfield  {author} {\bibinfo {author} {\bibfnamefont {T.}~\bibnamefont
  {Torres}}, \bibinfo {author} {\bibfnamefont {S.}~\bibnamefont {Patrick}},
  \bibinfo {author} {\bibfnamefont {A.}~\bibnamefont {Coutant}}, \bibinfo
  {author} {\bibfnamefont {M.}~\bibnamefont {Richartz}}, \bibinfo {author}
  {\bibfnamefont {E.~W.}\ \bibnamefont {Tedford}}, \ and\ \bibinfo {author}
  {\bibfnamefont {S.}~\bibnamefont {Weinfurtner}},\ }\href {\doibase
  10.1038/nphys4151} {\bibfield  {journal} {\bibinfo  {journal} {Nature Phys.}\
  }\textbf {\bibinfo {volume} {13}},\ \bibinfo {pages} {833} (\bibinfo {year}
  {2017}{\natexlab{a}})},\ \Eprint {http://arxiv.org/abs/1612.06180}
  {arXiv:1612.06180 [gr-qc]} \BibitemShut {NoStop}%
\bibitem [{\citenamefont {Vocke}\ \emph {et~al.}(2017)\citenamefont {Vocke},
  \citenamefont {Maitland}, \citenamefont {Prain}, \citenamefont {Biancalana},
  \citenamefont {Marino},\ and\ \citenamefont {Faccio}}]{prain_exp}%
  \BibitemOpen
  \bibfield  {author} {\bibinfo {author} {\bibfnamefont {D.}~\bibnamefont
  {Vocke}}, \bibinfo {author} {\bibfnamefont {C.}~\bibnamefont {Maitland}},
  \bibinfo {author} {\bibfnamefont {A.}~\bibnamefont {Prain}}, \bibinfo
  {author} {\bibfnamefont {F.}~\bibnamefont {Biancalana}}, \bibinfo {author}
  {\bibfnamefont {F.}~\bibnamefont {Marino}}, \ and\ \bibinfo {author}
  {\bibfnamefont {D.}~\bibnamefont {Faccio}},\ }\href@noop {} {\bibfield
  {journal} {\bibinfo  {journal} {arXiv preprint arXiv:1709.04293}\ } (\bibinfo
  {year} {2017})}\BibitemShut {NoStop}%
\bibitem [{\citenamefont {Stepanyants}\ and\ \citenamefont
  {Yeoh}(2008)}]{stepanyants}%
  \BibitemOpen
  \bibfield  {author} {\bibinfo {author} {\bibfnamefont {Y.~A.}\ \bibnamefont
  {Stepanyants}}\ and\ \bibinfo {author} {\bibfnamefont {G.~H.}\ \bibnamefont
  {Yeoh}},\ }\href@noop {} {\bibfield  {journal} {\bibinfo  {journal} {Journal
  of Fluid Mechanics}\ }\textbf {\bibinfo {volume} {604}},\ \bibinfo {pages}
  {77} (\bibinfo {year} {2008})}\BibitemShut {NoStop}%
\bibitem [{\citenamefont {Andersen}\ \emph {et~al.}(2003)\citenamefont
  {Andersen}, \citenamefont {Bohr}, \citenamefont {Stenum}, \citenamefont
  {Rasmussen},\ and\ \citenamefont {Lautrup}}]{andersen}%
  \BibitemOpen
  \bibfield  {author} {\bibinfo {author} {\bibfnamefont {A.}~\bibnamefont
  {Andersen}}, \bibinfo {author} {\bibfnamefont {T.}~\bibnamefont {Bohr}},
  \bibinfo {author} {\bibfnamefont {B.}~\bibnamefont {Stenum}}, \bibinfo
  {author} {\bibfnamefont {J.~J.}\ \bibnamefont {Rasmussen}}, \ and\ \bibinfo
  {author} {\bibfnamefont {B.}~\bibnamefont {Lautrup}},\ }\href@noop {}
  {\bibfield  {journal} {\bibinfo  {journal} {Physical review letters}\
  }\textbf {\bibinfo {volume} {91}},\ \bibinfo {pages} {104502} (\bibinfo
  {year} {2003})}\BibitemShut {NoStop}%
\bibitem [{\citenamefont {Cardoso}\ \emph {et~al.}(2013)\citenamefont
  {Cardoso}, \citenamefont {Carucci}, \citenamefont {Pani},\ and\ \citenamefont
  {Sotiriou}}]{Thomas1}%
  \BibitemOpen
  \bibfield  {author} {\bibinfo {author} {\bibfnamefont {V.}~\bibnamefont
  {Cardoso}}, \bibinfo {author} {\bibfnamefont {I.~P.}\ \bibnamefont
  {Carucci}}, \bibinfo {author} {\bibfnamefont {P.}~\bibnamefont {Pani}}, \
  and\ \bibinfo {author} {\bibfnamefont {T.~P.}\ \bibnamefont {Sotiriou}},\
  }\href {\doibase 10.1103/PhysRevLett.111.111101} {\bibfield  {journal}
  {\bibinfo  {journal} {Phys. Rev. Lett.}\ }\textbf {\bibinfo {volume} {111}},\
  \bibinfo {pages} {111101} (\bibinfo {year} {2013})},\ \Eprint
  {http://arxiv.org/abs/1308.6587} {arXiv:1308.6587 [gr-qc]} \BibitemShut
  {NoStop}%
\bibitem [{\citenamefont {Coates}\ \emph {et~al.}(2017)\citenamefont {Coates},
  \citenamefont {Horbartsch},\ and\ \citenamefont {Sotiriou}}]{Thomas2}%
  \BibitemOpen
  \bibfield  {author} {\bibinfo {author} {\bibfnamefont {A.}~\bibnamefont
  {Coates}}, \bibinfo {author} {\bibfnamefont {M.~W.}\ \bibnamefont
  {Horbartsch}}, \ and\ \bibinfo {author} {\bibfnamefont {T.~P.}\ \bibnamefont
  {Sotiriou}},\ }\href {\doibase 10.1103/PhysRevD.95.084003} {\bibfield
  {journal} {\bibinfo  {journal} {Phys. Rev.}\ }\textbf {\bibinfo {volume} {D
  95}},\ \bibinfo {pages} {084003} (\bibinfo {year} {2017})},\ \Eprint
  {http://arxiv.org/abs/1606.03981} {arXiv:1606.03981 [gr-qc]} \BibitemShut
  {NoStop}%
\bibitem [{\citenamefont {Hod}(2015)}]{QBS_kerr}%
  \BibitemOpen
  \bibfield  {author} {\bibinfo {author} {\bibfnamefont {S.}~\bibnamefont
  {Hod}},\ }\href@noop {} {\bibfield  {journal} {\bibinfo  {journal} {Physics
  Letters B}\ }\textbf {\bibinfo {volume} {749}},\ \bibinfo {pages} {167}
  (\bibinfo {year} {2015})}\BibitemShut {NoStop}%
\bibitem [{\citenamefont {Dolan}\ and\ \citenamefont
  {Dempsey}(2015)}]{dolandirac}%
  \BibitemOpen
  \bibfield  {author} {\bibinfo {author} {\bibfnamefont {S.~R.}\ \bibnamefont
  {Dolan}}\ and\ \bibinfo {author} {\bibfnamefont {D.}~\bibnamefont
  {Dempsey}},\ }\href@noop {} {\bibfield  {journal} {\bibinfo  {journal}
  {Classical and Quantum Gravity}\ }\textbf {\bibinfo {volume} {32}},\ \bibinfo
  {pages} {184001} (\bibinfo {year} {2015})}\BibitemShut {NoStop}%
\bibitem [{\citenamefont {Barausse}\ \emph {et~al.}(2014)\citenamefont
  {Barausse}, \citenamefont {Cardoso},\ and\ \citenamefont
  {Pani}}]{environment}%
  \BibitemOpen
  \bibfield  {author} {\bibinfo {author} {\bibfnamefont {E.}~\bibnamefont
  {Barausse}}, \bibinfo {author} {\bibfnamefont {V.}~\bibnamefont {Cardoso}}, \
  and\ \bibinfo {author} {\bibfnamefont {P.}~\bibnamefont {Pani}},\ }\href@noop
  {} {\bibfield  {journal} {\bibinfo  {journal} {Physical Review D}\ }\textbf
  {\bibinfo {volume} {89}},\ \bibinfo {pages} {104059} (\bibinfo {year}
  {2014})}\BibitemShut {NoStop}%
\bibitem [{\citenamefont {Cardoso}\ \emph {et~al.}()\citenamefont {Cardoso},
  \citenamefont {Lemos},\ and\ \citenamefont {Yoshida}}]{QNM_DBT_stab}%
  \BibitemOpen
  \bibfield  {author} {\bibinfo {author} {\bibfnamefont {V.}~\bibnamefont
  {Cardoso}}, \bibinfo {author} {\bibfnamefont {J.}~\bibnamefont {Lemos}}, \
  and\ \bibinfo {author} {\bibfnamefont {S.}~\bibnamefont {Yoshida}},\
  }\href@noop {} {\bibfield  {journal} {\bibinfo  {journal} {arXiv preprint
  gr-qc/0410107}\ }\textbf {\bibinfo {volume} {3}},\ \bibinfo {pages}
  {5}}\BibitemShut {NoStop}%
\bibitem [{\citenamefont {Berti}\ \emph {et~al.}(2004)\citenamefont {Berti},
  \citenamefont {Cardoso},\ and\ \citenamefont {Lemos}}]{QNM_DBT}%
  \BibitemOpen
  \bibfield  {author} {\bibinfo {author} {\bibfnamefont {E.}~\bibnamefont
  {Berti}}, \bibinfo {author} {\bibfnamefont {V.}~\bibnamefont {Cardoso}}, \
  and\ \bibinfo {author} {\bibfnamefont {J.~P.}\ \bibnamefont {Lemos}},\
  }\href@noop {} {\bibfield  {journal} {\bibinfo  {journal} {Physical Review
  D}\ }\textbf {\bibinfo {volume} {70}},\ \bibinfo {pages} {124006} (\bibinfo
  {year} {2004})}\BibitemShut {NoStop}%
\bibitem [{\citenamefont {Dolan}\ \emph {et~al.}(2012)\citenamefont {Dolan},
  \citenamefont {Oliveira},\ and\ \citenamefont {Crispino}}]{DBT_resonances}%
  \BibitemOpen
  \bibfield  {author} {\bibinfo {author} {\bibfnamefont {S.~R.}\ \bibnamefont
  {Dolan}}, \bibinfo {author} {\bibfnamefont {L.~A.}\ \bibnamefont {Oliveira}},
  \ and\ \bibinfo {author} {\bibfnamefont {L.~C.}\ \bibnamefont {Crispino}},\
  }\href@noop {} {\bibfield  {journal} {\bibinfo  {journal} {Physical Review
  D}\ }\textbf {\bibinfo {volume} {85}},\ \bibinfo {pages} {044031} (\bibinfo
  {year} {2012})}\BibitemShut {NoStop}%
\bibitem [{\citenamefont {Lepe}\ and\ \citenamefont
  {Saavedra}(2005)}]{lepe2005}%
  \BibitemOpen
  \bibfield  {author} {\bibinfo {author} {\bibfnamefont {S.}~\bibnamefont
  {Lepe}}\ and\ \bibinfo {author} {\bibfnamefont {J.}~\bibnamefont
  {Saavedra}},\ }\href@noop {} {\bibfield  {journal} {\bibinfo  {journal}
  {Physics Letters B}\ }\textbf {\bibinfo {volume} {617}},\ \bibinfo {pages}
  {174} (\bibinfo {year} {2005})}\BibitemShut {NoStop}%
\bibitem [{\citenamefont {Lautrup}(2005)}]{lautrup}%
  \BibitemOpen
  \bibfield  {author} {\bibinfo {author} {\bibfnamefont {B.}~\bibnamefont
  {Lautrup}},\ }\href@noop {} {\bibfield  {journal} {\bibinfo  {journal}
  {Exotic and Everyday Phenomena in the Macroscopic World, IOP}\ } (\bibinfo
  {year} {2005})}\BibitemShut {NoStop}%
\bibitem [{\citenamefont {Rosenhead}(1930)}]{rosenhead}%
  \BibitemOpen
  \bibfield  {author} {\bibinfo {author} {\bibfnamefont {L.}~\bibnamefont
  {Rosenhead}},\ }\href@noop {} {\bibfield  {journal} {\bibinfo  {journal}
  {Proceedings of the Royal Society of London. Series A, Containing papers of a
  mathematical and physical character}\ }\textbf {\bibinfo {volume} {127}},\
  \bibinfo {pages} {590} (\bibinfo {year} {1930})}\BibitemShut {NoStop}%
\bibitem [{\citenamefont {Mih}(1990)}]{mih2}%
  \BibitemOpen
  \bibfield  {author} {\bibinfo {author} {\bibfnamefont {W.~C.}\ \bibnamefont
  {Mih}},\ }\href {\doibase 10.1080/00221689009499078} {\bibfield  {journal}
  {\bibinfo  {journal} {Journal of Hydraulic Research}\ }\textbf {\bibinfo
  {volume} {28}},\ \bibinfo {pages} {392} (\bibinfo {year} {1990})}\BibitemShut
  {NoStop}%
\bibitem [{\citenamefont {Vatistas}(1989)}]{vatistas}%
  \BibitemOpen
  \bibfield  {author} {\bibinfo {author} {\bibfnamefont {G.}~\bibnamefont
  {Vatistas}},\ }\href@noop {} {\bibfield  {journal} {\bibinfo  {journal}
  {Journal of Hydraulic Research}\ }\textbf {\bibinfo {volume} {27}},\ \bibinfo
  {pages} {417} (\bibinfo {year} {1989})}\BibitemShut {NoStop}%
\bibitem [{\citenamefont {Hite~Jr}\ and\ \citenamefont
  {Mih}(1994)}]{hydraulic}%
  \BibitemOpen
  \bibfield  {author} {\bibinfo {author} {\bibfnamefont {J.~E.}\ \bibnamefont
  {Hite~Jr}}\ and\ \bibinfo {author} {\bibfnamefont {W.~C.}\ \bibnamefont
  {Mih}},\ }\href@noop {} {\bibfield  {journal} {\bibinfo  {journal} {Journal
  of hydraulic Engineering}\ }\textbf {\bibinfo {volume} {120}},\ \bibinfo
  {pages} {284} (\bibinfo {year} {1994})}\BibitemShut {NoStop}%
\bibitem [{\citenamefont {B{\"u}hler}(2014)}]{buhble}%
  \BibitemOpen
  \bibfield  {author} {\bibinfo {author} {\bibfnamefont {O.}~\bibnamefont
  {B{\"u}hler}},\ }\href@noop {} {\emph {\bibinfo {title} {Waves and mean
  flows}}}\ (\bibinfo  {publisher} {Cambridge University Press},\ \bibinfo
  {year} {2014})\BibitemShut {NoStop}%
\bibitem [{\citenamefont {Sch{\"u}tzhold}\ and\ \citenamefont
  {Unruh}(2002{\natexlab{b}})}]{gravwaves}%
  \BibitemOpen
  \bibfield  {author} {\bibinfo {author} {\bibfnamefont {R.}~\bibnamefont
  {Sch{\"u}tzhold}}\ and\ \bibinfo {author} {\bibfnamefont {W.~G.}\
  \bibnamefont {Unruh}},\ }\href@noop {} {\bibfield  {journal} {\bibinfo
  {journal} {Physical Review D}\ }\textbf {\bibinfo {volume} {66}},\ \bibinfo
  {pages} {044019} (\bibinfo {year} {2002}{\natexlab{b}})}\BibitemShut
  {NoStop}%
\bibitem [{\citenamefont {Kopiev}\ and\ \citenamefont
  {Belyaev}(2010)}]{kopiev}%
  \BibitemOpen
  \bibfield  {author} {\bibinfo {author} {\bibfnamefont {V.~F.}\ \bibnamefont
  {Kopiev}}\ and\ \bibinfo {author} {\bibfnamefont {I.~V.}\ \bibnamefont
  {Belyaev}},\ }\href@noop {} {\bibfield  {journal} {\bibinfo  {journal}
  {Journal of Sound and Vibration}\ }\textbf {\bibinfo {volume} {329}},\
  \bibinfo {pages} {1409} (\bibinfo {year} {2010})}\BibitemShut {NoStop}%
\bibitem [{\citenamefont {Coste}\ \emph {et~al.}(1999)\citenamefont {Coste},
  \citenamefont {Lund},\ and\ \citenamefont {Umeki}}]{coste}%
  \BibitemOpen
  \bibfield  {author} {\bibinfo {author} {\bibfnamefont {C.}~\bibnamefont
  {Coste}}, \bibinfo {author} {\bibfnamefont {F.}~\bibnamefont {Lund}}, \ and\
  \bibinfo {author} {\bibfnamefont {M.}~\bibnamefont {Umeki}},\ }\href@noop {}
  {\bibfield  {journal} {\bibinfo  {journal} {Physical Review E}\ }\textbf
  {\bibinfo {volume} {60}},\ \bibinfo {pages} {4908} (\bibinfo {year}
  {1999})}\BibitemShut {NoStop}%
\bibitem [{\citenamefont {Cardoso}\ \emph {et~al.}(2004)\citenamefont
  {Cardoso}, \citenamefont {Lemos},\ and\ \citenamefont {Yoshida}}]{Cardoso04}%
  \BibitemOpen
  \bibfield  {author} {\bibinfo {author} {\bibfnamefont {V.}~\bibnamefont
  {Cardoso}}, \bibinfo {author} {\bibfnamefont {J.~P.~S.}\ \bibnamefont
  {Lemos}}, \ and\ \bibinfo {author} {\bibfnamefont {S.}~\bibnamefont
  {Yoshida}},\ }\href {\doibase 10.1103/PhysRevD.70.124032} {\bibfield
  {journal} {\bibinfo  {journal} {Phys. Rev.}\ }\textbf {\bibinfo {volume} {D
  70}},\ \bibinfo {pages} {124032} (\bibinfo {year} {2004})},\ \Eprint
  {http://arxiv.org/abs/gr-qc/0410107} {arXiv:gr-qc/0410107 [gr-qc]}
  \BibitemShut {NoStop}%
\bibitem [{\citenamefont {Torres}\ \emph
  {et~al.}(2017{\natexlab{b}})\citenamefont {Torres}, \citenamefont {Coutant},
  \citenamefont {Dolan},\ and\ \citenamefont {Weinfurtner}}]{Awesome_paper}%
  \BibitemOpen
  \bibfield  {author} {\bibinfo {author} {\bibfnamefont {T.}~\bibnamefont
  {Torres}}, \bibinfo {author} {\bibfnamefont {A.}~\bibnamefont {Coutant}},
  \bibinfo {author} {\bibfnamefont {S.}~\bibnamefont {Dolan}}, \ and\ \bibinfo
  {author} {\bibfnamefont {S.}~\bibnamefont {Weinfurtner}},\ }\href@noop {} {\
  (\bibinfo {year} {2017}{\natexlab{b}})},\ \Eprint
  {http://arxiv.org/abs/1712.04675} {arXiv:1712.04675 [gr-qc]} \BibitemShut
  {NoStop}%
\bibitem [{\citenamefont {Cardoso}\ \emph {et~al.}(2009)\citenamefont
  {Cardoso}, \citenamefont {Miranda}, \citenamefont {Berti}, \citenamefont
  {Witek},\ and\ \citenamefont {Zanchin}}]{CardosoLR}%
  \BibitemOpen
  \bibfield  {author} {\bibinfo {author} {\bibfnamefont {V.}~\bibnamefont
  {Cardoso}}, \bibinfo {author} {\bibfnamefont {A.~S.}\ \bibnamefont
  {Miranda}}, \bibinfo {author} {\bibfnamefont {E.}~\bibnamefont {Berti}},
  \bibinfo {author} {\bibfnamefont {H.}~\bibnamefont {Witek}}, \ and\ \bibinfo
  {author} {\bibfnamefont {V.~T.}\ \bibnamefont {Zanchin}},\ }\href@noop {}
  {\bibfield  {journal} {\bibinfo  {journal} {Phys. Rev.}\ }\textbf {\bibinfo
  {volume} {D 79}},\ \bibinfo {pages} {064016} (\bibinfo {year}
  {2009})}\BibitemShut {NoStop}%
\bibitem [{Note1()}]{Note1}%
  \BibitemOpen
  \bibinfo {note} {See Supplemental Material, which includes Refs.~\cite
  {berrymount,Froman,CoutantCF,leaverkerr,leaverrn,nollert,Cardoso04,gautschi,Cook14,Konoplya:2011qq},
  for a detailed description of the WKB method, the continued fraction method,
  and our numerical simulations, including a comparison between
  them.}\BibitemShut {Stop}%
\bibitem [{\citenamefont {Leaver}(1985)}]{leaverkerr}%
  \BibitemOpen
  \bibfield  {author} {\bibinfo {author} {\bibfnamefont {E.~W.}\ \bibnamefont
  {Leaver}},\ }\href {\doibase 10.1098/rspa.1985.0119} {\bibfield  {journal}
  {\bibinfo  {journal} {Proc. Roy. Soc. Lond.}\ }\textbf {\bibinfo {volume} {A
  402}},\ \bibinfo {pages} {285} (\bibinfo {year} {1985})}\BibitemShut
  {NoStop}%
\bibitem [{\citenamefont {Nollert}(1993)}]{nollert}%
  \BibitemOpen
  \bibfield  {author} {\bibinfo {author} {\bibfnamefont {H.-P.}\ \bibnamefont
  {Nollert}},\ }\href {\doibase 10.1103/PhysRevD.47.5253} {\bibfield  {journal}
  {\bibinfo  {journal} {Phys. Rev. D}\ }\textbf {\bibinfo {volume} {47}},\
  \bibinfo {pages} {5253} (\bibinfo {year} {1993})}\BibitemShut {NoStop}%
\bibitem [{\citenamefont {Benone}\ \emph {et~al.}(2015)\citenamefont {Benone},
  \citenamefont {Crispino}, \citenamefont {Herdeiro},\ and\ \citenamefont
  {Radu}}]{Benone14}%
  \BibitemOpen
  \bibfield  {author} {\bibinfo {author} {\bibfnamefont {C.~L.}\ \bibnamefont
  {Benone}}, \bibinfo {author} {\bibfnamefont {L.~C.~B.}\ \bibnamefont
  {Crispino}}, \bibinfo {author} {\bibfnamefont {C.}~\bibnamefont {Herdeiro}},
  \ and\ \bibinfo {author} {\bibfnamefont {E.}~\bibnamefont {Radu}},\ }\href
  {\doibase 10.1103/PhysRevD.91.104038} {\bibfield  {journal} {\bibinfo
  {journal} {Phys. Rev.}\ }\textbf {\bibinfo {volume} {D 91}},\ \bibinfo
  {pages} {104038} (\bibinfo {year} {2015})},\ \Eprint
  {http://arxiv.org/abs/1412.7278} {arXiv:1412.7278 [gr-qc]} \BibitemShut
  {NoStop}%
\bibitem [{\citenamefont {Berry}\ and\ \citenamefont
  {Mount}(1972)}]{berrymount}%
  \BibitemOpen
  \bibfield  {author} {\bibinfo {author} {\bibfnamefont {M.~V.}\ \bibnamefont
  {Berry}}\ and\ \bibinfo {author} {\bibfnamefont {K.}~\bibnamefont {Mount}},\
  }\href@noop {} {\bibfield  {journal} {\bibinfo  {journal} {Reports on
  Progress in Physics}\ }\textbf {\bibinfo {volume} {35}},\ \bibinfo {pages}
  {315} (\bibinfo {year} {1972})}\BibitemShut {NoStop}%
\bibitem [{\citenamefont {Fr{\"o}man}\ and\ \citenamefont
  {Fr{\"o}man}(2002)}]{Froman}%
  \BibitemOpen
  \bibfield  {author} {\bibinfo {author} {\bibfnamefont {N.}~\bibnamefont
  {Fr{\"o}man}}\ and\ \bibinfo {author} {\bibfnamefont {P.~O.}\ \bibnamefont
  {Fr{\"o}man}},\ }\href@noop {} {\emph {\bibinfo {title} {Physical problems
  solved by the phase-integral method}}}\ (\bibinfo  {publisher} {Cambridge
  University Press},\ \bibinfo {year} {2002})\BibitemShut {NoStop}%
\bibitem [{\citenamefont {Coutant}\ \emph {et~al.}(2012)\citenamefont
  {Coutant}, \citenamefont {Parentani},\ and\ \citenamefont
  {Finazzi}}]{CoutantCF}%
  \BibitemOpen
  \bibfield  {author} {\bibinfo {author} {\bibfnamefont {A.}~\bibnamefont
  {Coutant}}, \bibinfo {author} {\bibfnamefont {R.}~\bibnamefont {Parentani}},
  \ and\ \bibinfo {author} {\bibfnamefont {S.}~\bibnamefont {Finazzi}},\ }\href
  {\doibase 10.1103/PhysRevD.85.024021} {\bibfield  {journal} {\bibinfo
  {journal} {Phys. Rev.}\ }\textbf {\bibinfo {volume} {D 85}},\ \bibinfo
  {pages} {024021} (\bibinfo {year} {2012})},\ \Eprint
  {http://arxiv.org/abs/1108.1821} {arXiv:1108.1821 [hep-th]} \BibitemShut
  {NoStop}%
\bibitem [{\citenamefont {Leaver}(1990)}]{leaverrn}%
  \BibitemOpen
  \bibfield  {author} {\bibinfo {author} {\bibfnamefont {E.~W.}\ \bibnamefont
  {Leaver}},\ }\href@noop {} {\bibfield  {journal} {\bibinfo  {journal}
  {Physical Review D}\ }\textbf {\bibinfo {volume} {41}},\ \bibinfo {pages}
  {2986} (\bibinfo {year} {1990})}\BibitemShut {NoStop}%
\bibitem [{\citenamefont {Gautschi}(1967)}]{gautschi}%
  \BibitemOpen
  \bibfield  {author} {\bibinfo {author} {\bibfnamefont {W.}~\bibnamefont
  {Gautschi}},\ }\href@noop {} {\bibfield  {journal} {\bibinfo  {journal} {SIAM
  review}\ }\textbf {\bibinfo {volume} {9}},\ \bibinfo {pages} {24} (\bibinfo
  {year} {1967})}\BibitemShut {NoStop}%
\bibitem [{\citenamefont {Cook}\ and\ \citenamefont
  {Zalutskiy}(2014)}]{Cook14}%
  \BibitemOpen
  \bibfield  {author} {\bibinfo {author} {\bibfnamefont {G.~B.}\ \bibnamefont
  {Cook}}\ and\ \bibinfo {author} {\bibfnamefont {M.}~\bibnamefont
  {Zalutskiy}},\ }\href {\doibase 10.1103/PhysRevD.90.124021} {\bibfield
  {journal} {\bibinfo  {journal} {Phys. Rev.}\ }\textbf {\bibinfo {volume} {D
  90}},\ \bibinfo {pages} {124021} (\bibinfo {year} {2014})},\ \Eprint
  {http://arxiv.org/abs/1410.7698} {arXiv:1410.7698 [gr-qc]} \BibitemShut
  {NoStop}%
\end{thebibliography}%

\newpage
\appendix
\begin{center}
\textbf{\large Supplemental Material}
\end{center}

\section{Example of vortex spirals} \label{mol}

\begin{figure} [H]
\begin{center}
\includegraphics[width=\linewidth]{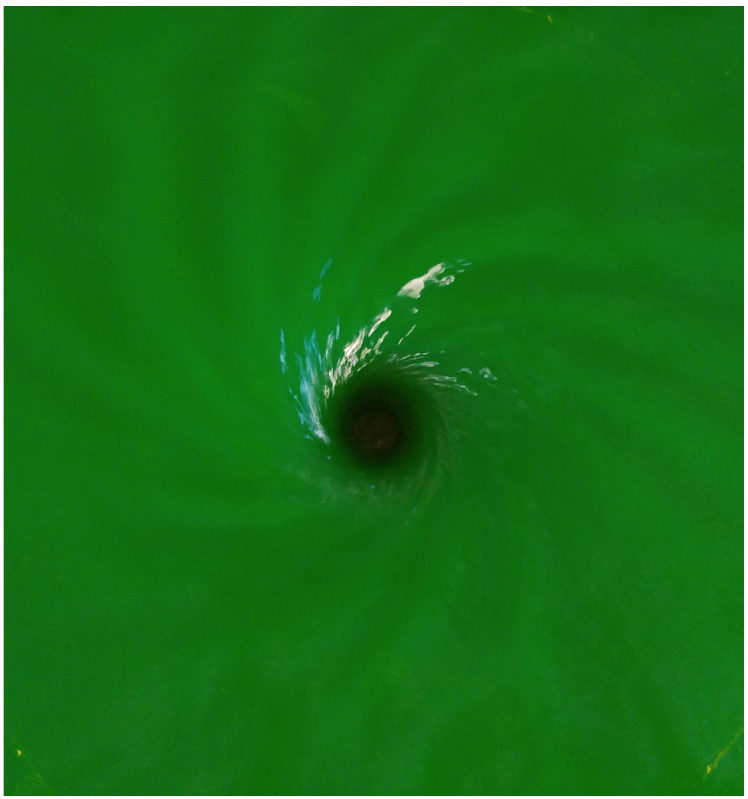} 
\caption{Example of the spirals that can be observed close to the drain of a perturbed DBT vortex. Like the dominant mode we observe in our MOL simulations the spiral is counter propagating, hinting that resonant frequencies could provide a good description of this phenomenon. This image is taken from a vortex flow obtained using the experimental setup~\cite{Awesome_paper}. 
Here the  water height is $5$cm at the outer boundaries of the basin and the water is drained at a rate of $30$ l/min through a hole of $4\mathrm{cm}$ diameter. Although the presented flow is too deep to be accurately described by the two-dimensional equations (as in the text), we believe that the observed spiral is qualitatively similar to those studied in this paper.} \label{fig:vtx}
\end{center}
\end{figure}

\section{WKB formula for QBS frequencies} \label{QBSlifetime}

The complex frequency $\omega_n$ of a QBS can be approximated by performing a tunnelling amplitude calculation using semi-classical wave mechanics~\cite{berrymount}. 
A solution to the wave equation can be approximated as a WKB mode everywhere except at the turning points of the scattering potential $(r_{*j})_{j=1...4}$ where $V(r_{*j})=0$ (see Fig.~1). To find solutions near the turning point, one can make the expansion $V(r_*) \simeq V(r_{*j}) + V'(r_{*j})(r_*-r_{*j})$. Eq.~(6) in the text becomes the Airy equation $-\partial_{r_*}^2\phi_{\omega m} + V'(r_{*j})(r-r_{*j})\phi_{\omega m} = 0$ whose solutions can be expressed in terms of the orthogonal functions $\mathrm{Ai}(r_*)$ and $\mathrm{Bi}(r_*)$. These take the asymptotic forms,
\begin{equation} \label{airy}
\begin{split}
\mathrm{Ai}(z) & \underset{-\infty}\sim \frac{1}{2|z|^{1/4}\sqrt{\pi}}\Big(e^{-i\frac{2}{3}(-z)^{3/2}+i\frac{\pi}{4}}+e^{i\frac{2}{3}(-z)^{3/2}-i\frac{\pi}{4}}\Big), \\
& \underset{\infty}\sim \frac{e^{-\frac{2}{3}z^{3/2}}}{2|z|^{1/4}\sqrt{\pi}}, \\
\mathrm{Bi}(z) & \underset{-\infty}\sim \frac{i}{2|z|^{1/4}\sqrt{\pi}}\Big(e^{i\frac{2}{3}(z)^{3/2}-i\frac{\pi}{4}}-e^{-i\frac{2}{3}(-z)^{3/2}+i\frac{\pi}{4}}\Big), \\
& \underset{\infty}\sim \frac{e^{\frac{2}{3}z^{3/2}}}{2
|z|^{1/4}\sqrt{\pi}},\\
\end{split}
\end{equation}
\noindent where $z = V'(r_{*j})^{1/3}(r_*-r_{*j})$, with $z<0$ defined as the classically allowed region and $z>0$ the classically forbidden region. A global solution of Eq.~(4) in the text is constructed by relating WKB modes $\phi_{\omega m} = \exp(\pm i\int k dr_*)/\sqrt{4\pi k}$ where $V$ is approximately linear, to Airy functions of large arguments through Eq.~\eqref{airy}. Let $\phi^{\rightarrow}_{\omega m}$ ($\phi^{\leftarrow}_{\omega m}$) be the right (left) moving mode in the region $z<0$ and $\phi^{\uparrow}_{\omega m}$ ($\phi^{\downarrow}_{\omega m}$) the growing (decaying) mode in the region $z>0$. Close to the turning point where $V(r_*)\simeq V'(r_{*j})(r_*-r_{*j})$, these modes take the form,
\begin{equation}
\begin{split}
\phi^{\rightarrow}_{\omega m}\simeq\frac{e^{-i\frac{2}{3}(-z)^{3/2}}}{2|z|^{1/4}\sqrt{\pi}}, \qquad \phi^{\leftarrow}_{\omega m}\simeq\frac{e^{i\frac{2}{3}(-z)^{3/2}}}{2|z|^{1/4}\sqrt{\pi}}, \\
\phi^{\uparrow}_{\omega m}\simeq\frac{e^{\frac{2}{3}z^{3/2}}}{2|z|^{1/4}\sqrt{\pi}}, \qquad \phi^{\downarrow}_{\omega m}\simeq\frac{e^{-\frac{2}{3}z^{3/2}}}{2|z|^{1/4}\sqrt{\pi}}. 
\end{split}
\end{equation}
\noindent Now, let $\phi_{\omega m}(z)$ be the globally defined solution which satisfies,
\begin{equation} \label{CF}
\begin{split}
\phi_{\omega m}(z<0) = & A^{\rightarrow}\phi^{\rightarrow}_{\omega m}(z)+A^{\leftarrow}\phi^{\leftarrow}_{\omega m}(z), \\
\phi_{\omega m}(z>0) = &  A^{\downarrow}\phi^{\downarrow}_{\omega m}(z)+A^{\uparrow}\phi^{\uparrow}_{\omega m}(z).
\end{split}
\end{equation}
\noindent Relating the two solutions to one another through Eq.~\eqref{airy} we see that,
\begin{equation} \label{1tp_cf}
\begin{pmatrix}
A^{\rightarrow}\\
A^{\leftarrow}
\end{pmatrix}
= T\cdot \begin{pmatrix}
A^{\downarrow}\\
A^{\uparrow}
\end{pmatrix}, \qquad T = e^{i\frac{\pi}{4}}\begin{pmatrix}
1 & -i/2\\
-i & 1/2
\end{pmatrix},
\end{equation}
\noindent where $T$ is the transfer matrix relating exponential modes to the oscillatory modes. Eq.~\eqref{1tp_cf} is the well-known single-turning-point connection formula~\cite{berrymount,Froman}. 
Notice that there is an ambiguity in choosing the second solution of the Airy equation (this is the standard irreversibility of connection formulae issue~\cite{Froman}). 
Indeed, any combination of $\mathrm{Ai}$ and $\mathrm{Bi}$ with a non-zero coefficient for $\mathrm{Bi}$ will have the same asymptotics as in Eq.~\eqref{airy}, and that would lead to a different connection formula. That ambiguity is lifted by requiring that the second solution has a vanishing Wronskian (defined as $(\phi_1|\phi_2) = \phi_1^* \phi_2' - \phi_1'^* \phi_2$). This defines $\mathrm{Bi}$, and ensures that the connection formula \eqref{CF} preserves the Wronskian conservation~\cite{CoutantCF}. 
In other words, $T \in U(1,1)$. The complex conjugate $\tilde{T}=T^*$ relates exponential modes to oscillatory modes in the mirror situation. The inverses $T^{-1}$ and $\tilde{T}^{-1}$ relate oscillatory to exponential modes.

This can now be applied to the scattering problem for the potential in Eq.~(7) in the text. To do so we combine the (single turning point) connection formula \eqref{CF} with WKB propagation in between. For this, in each region where WKB is valid, we decompose the solution on a WKB basis as $\phi_{\omega m} = A_+ \exp(+ i\int k dr_*)/\sqrt{4\pi k} + A_- \exp(- i\int k dr_*)/\sqrt{4\pi k}$ and relate the amplitudes in each regions using the connection formula \eqref{1tp_cf}. Then, the amplitudes of plane waves on both asymptotics $r_* \to \pm \infty$ are related by, 
\begin{equation} \label{scattering}
\begin{pmatrix}
A_{\mathrm{H}}\\
0
\end{pmatrix} = TJ_1\tilde{T}^{-1}J_dTJ_2\tilde{T}^{-1} \begin{pmatrix}
A_{\mathrm{in}}\\
A_{\infty}
\end{pmatrix},
\end{equation}
\noindent where the $J$ matrices are the propagation matrices of the exponential modes underneath the peaks and the oscillatory modes in the dip, i.e.
\begin{equation}
J_d = \begin{pmatrix}
e^{iS_d} & 0 \\
0 & e^{-iS_d}
\end{pmatrix}, \qquad J_{1,2} = \begin{pmatrix}
0 & e^{S_{1,2}} \\
e^{-S_{1,2}} & 0
\end{pmatrix},
\end{equation}
\noindent where $S_{d,1,2}=\int_{I_{d,1,2}} |k(r_*)|dr_*$ as before. Note that $J_{1,2}$ is anti-diagonal owing to the fact that a growing mode in one direction is a decaying mode in the other direction. However by looking at the boundary conditions in Eq.~(8) in the text, we see that if $\omega^\mathrm{R}<m\Omega_{\mathrm{H}}$ the mode will appear to propagate in the opposite direction and therefore $A_{\mathrm{H}}$ will swap places with the $0$ on the left of Eq.~\eqref{scattering}. QBSs are defined by the condition $A_{\mathrm{in}}=0$. Solving Eq.~\eqref{scattering} for this condition, one finds that,
\begin{equation} \label{cond}
e^{-2 i S_d} + R_2^* R_1 = 0,
\end{equation}
\noindent where the $R_{1,2}$'s are the reflection coefficients across the potential barriers, given by 
\be
R_{1,2} = e^{-i\frac{\pi}{2}}\Bigg(\frac{1-\frac{1}{4}e^{-2S_{1,2}}}{1+\frac{1}{4}e^{-2S_{1,2}}}\Bigg)^{\pm1}. 
\ee
The sign $\pm$ acts only on $R_2$ and is $-$ if the coefficient is superradiant, and $+$ otherwise.
To extract $\omega_n = \omega^\mathrm{R}_n + i\Gamma_n$ from Eq.~\eqref{cond}, we assume the QBS modes have a long lifetime, i.e. $\Gamma_n \ll \omega^\mathrm{R}_n$. We can then expand the action as $S(\omega_n) = S(\omega^\mathrm{R}_n) + i\Gamma_n\partial_{\omega}S|_{\omega=\omega^\mathrm{R}_n}$. The real part then gives the Bohr-Sommerfeld condition, i.e. Eq.~(12) in the text. The imaginary part gives the life-time as,
\be
\Gamma_n = \frac{\log|R_1R_2|}{2\partial_{\omega}S_d|_{\omega=\omega^\mathrm{R}_n}}. 
\ee
Notice that $\partial_{\omega}S_d|_{\omega=\omega^\mathrm{R}_n}$ is the semiclassical time for a wave packet to go from one turning point to the other in the potential well. Moreover, in the semiclassical limit, one has $e^{-S_k} \ll 1$, and hence the log in the above equation can be expanded, leading to the expression used in the core of the text, see Eq.~(13).

\section{Numerical simulations in the time domain} \label{mol} 

The wave equation~(4) in the text is a partial differential equation (PDE) in $(r,\theta,t)$ for the field $\phi$. However, we can exploit the symmetry of the system to write $\phi(t,r,\theta)=\phi_m(t,r)\exp(im\theta)/\sqrt{r}$ in which case the wave equation becomes a second order PDE in $(r,t)$ for $\phi_m$. To solve this equation we use the Method of Lines (MOL). The first step is to discretise equation~(4) in the text by approximating the $r$-derivatives using 5-point finite difference (FD) stencils, hence converting the differential equation into a matrix equation. Boundary conditions (BCs) in $r$ are implemented at the edge of the FD matrix. The first BC is placed just inside the horizon and is left free which is achieved using a one-sided stencil. This is justified since the horizon acts as a one-way membrane, hence the value of the field inside the horizon cannot affect the exterior. The second BC is a hard wall placed far from the vortex centre. The wall is placed sufficiently far away that reflections do not enter the region of interest within the timescale of the simulations. To solve in time we use a fourth order Runge-Kutta method (RK4) with a gaussian pulse as initial condition, i.e.
\begin{equation}
\phi(r,\theta,t=0) = \frac{\mathcal{A}}{\sqrt{2\pi\sigma^2}}\exp\left(-\frac{(r\cos\theta-x_0)^2}{2\sigma^2}\right),
\end{equation}
which is centred on $x_0\sim 5\sigma$, where $\sigma$ gives the spread of the gaussian, and $\mathcal{A}$ is an arbitrary amplitude. The first time derivative is imposed such that the pulse propagates toward the vortex, i.e.
\begin{equation}
\partial_t\phi|_{t=0} = c\left(\frac{r\cos\theta-x_0}{\sigma^2}\right)\phi.
\end{equation}
The corresponding conditions on the $m$-components are calculated through,
\begin{equation}
\phi_m(r,t=0) = \frac{\sqrt{r}}{2\pi}\int^{2\pi}_0 \phi(r,\theta,t=0)e^{-im\theta}d\theta,
\end{equation}
and similarly for $\partial_t\phi_m|_{t=0}$. The wave equation is then used to compute $\phi_m$ at the next time step. Once the pulse has passed the vortex, $\phi_m$ is purely outgoing and the response of the system can be observed. 

The frequency of the response $\omega=\omega^\mathrm{R}+i\Gamma$ is extracted using the time Fourier transform of $\phi_m$, i.e. $\tilde{\phi}_m(r_1,\omega)=\mathcal{F}[\phi_m(r_1,t)]$, over a range $[t_1,t_2]$. If $\phi_m$ is of the form $\exp(i\omega t)$, then $\tilde{\phi}$ is given by,
\begin{equation} \label{ftsignal}
\tilde{\phi}(\omega) = \frac{1-e^{-[\Gamma-i(\omega-\omega^\mathrm{R})](t_2-t_1)}}{\Gamma-i(\omega-\omega^\mathrm{R})}.
\end{equation}
We fit this model to our data, using non-linear regression to obtain the values of $\omega^\mathrm{R}$ and $\Gamma$.

\section{Continued fraction method} \label{sec:cfm}

The continued fraction method, in the context of black hole perturbations, was developed by Leaver~\cite{leaverkerr,leaverrn} 
and improved by Nollert~\cite{nollert}. 
The procedure, in the present context, consists in writing the solution of (4) (in the text) as a Frobenius expansion
\be \label{cf1}
\phi =e^{-i\omega t + i m \theta}e^{i \frac{\omega}{c} r} r^{-\frac{1}{2}+i\frac{\omega}{c} r_H} \sum_{n=0}^{\infty}a_n \left(\frac{r-r_H}{r}\right)^n.
\ee
For~\eqref{cf1} to be a solution of the wave equation~(4) in the text, the coefficients $a_n=a_n(m,\omega)$ have to  satisfy a $12$-term recurrence relation. (For the DBT vortex, the corresponding recurrence relation has only four terms~\cite{Cardoso04}). 
By successive Gaussian eliminations, we can reduce the $12$-term recurrence relation to a $3$-term recurrence relation of the form
\be 
\alpha_n a_{n+1} + \beta_n a_n + \gamma_n a_{n-1} = 0,
\ee
where $\alpha_n$, $\beta_n$, and $\gamma_n$ are coefficients that depend on $\omega$ and $m$.

Using the change of variable $r \to r_*$, expression~\eqref{cf1} above is compatible with the boundary conditions given in (8) in the text as long as the infinite sum is everywhere convergent. The ratio test guarantees that the sum converges if 
\be
\left| \frac{r-r_h}{r} \right| < \lim_{n \rightarrow \infty} \left| \frac{a_n}{a_{n+1}} \right|.
\ee
From the recurrence relation, one can show that $|a_n/a_{n+1}|\rightarrow 1$ and, therefore, the sum converges for all $r \in [r_h,\infty)$, independently of the frequency $\omega$. To assess convergence at spatial infinity, however, we need a theorem due to Pincherle~\cite{gautschi,Cook14}. 
This theorem guarantees that convergence at spatial infinity only occurs when the recurrence coefficients satisfy the following equation,
\be   
\beta_0  - \frac{\alpha_0 \gamma_1}{\beta_1 -}\frac{\alpha_1\gamma_2}{\beta_2 -}\frac{\alpha_2\gamma_3}{\beta_3 -} \dots = 0, \label{cf2}
\ee
where the expression on the left-hand-side is a continued fraction, written in standard notation (this explains the naming of the method). For a given $m$ value, the equation above picks a discrete set of frequencies $\omega$ that guarantee convergence of~\eqref{cf1} and, therefore, correspond to quasinormal modes or quasibound states of the problem under analysis.    

The numerical implementation of the method follows closely the algorithm described in Secs.~III.~H and III.~I of Ref.~\cite{Konoplya:2011qq}. 
To solve \eqref{cf2}, we use a damped Newton's method with initial guess given by the WKB method estimates. In order to guarantee at least 6-digit precision for the all the calculated frequencies, we truncate the continued fraction in \eqref{cf2} at 25000 terms. \\

\section{Comparison of results obtained from WKB, CFM and MOL methods} \label{mol}
\begin{figure*} [!tp]
\begin{center}
\includegraphics[width=\linewidth]{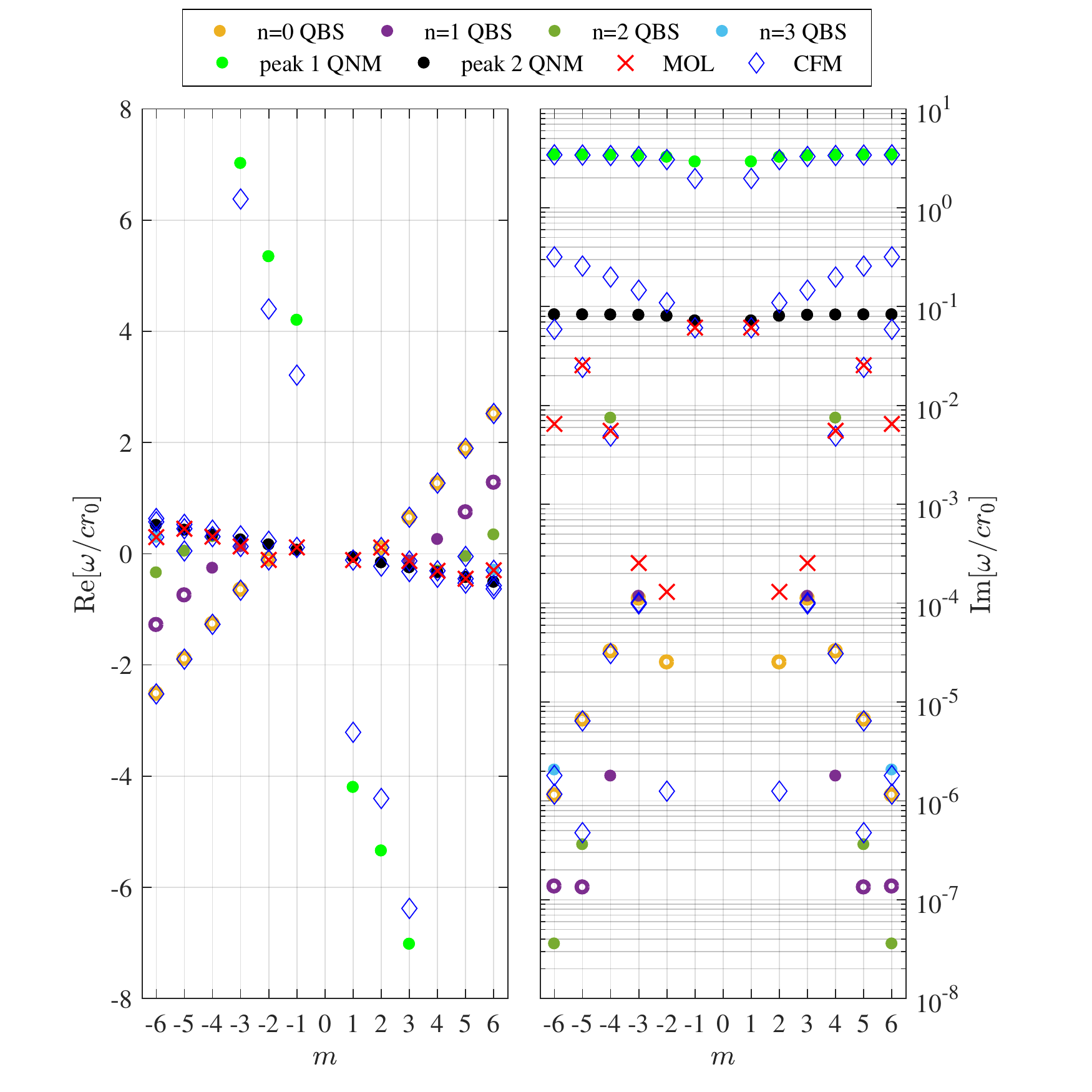} 
\caption{
The real (left panel) and imaginary (right panel) parts of QNM spectrum for the parameters $C/cr_0=3,D/cr_0=0.1$. Results are obtained by the three methods: MOL (red crosses), CFM (blue diamonds) and WKB (coloured dots for QBSs, bright green dots for QNMs on the inner potential barrier and black solid dots QNMs on the outer barrier). The different colours for the WKB points represent the different energy levels of the QBSs, with $n=0$ being the ground state in the potential well. Note, since on the right panel the results are displayed on a logarithmic scale, we plot the absolute value of the imaginary parts. There are some unstable modes (i.e.~$\Im(\omega)>0$), which are displayed as hollow points. These unstable modes were not seen in the time domain simulation, as their growth rate is extremely small. Under experimental conditions dissipative effects have to be taken into account, which may render these modes stable. This is subject to further investigation.}
 \label{fig:full_spectrum}
\end{center}
\end{figure*}

In Fig.~\ref{fig:full_spectrum}, we show the complex frequency spectrum for $|m| < 7$, but displaying all QBSs and QNMs (no overtones) even those that weren't excited in the time domain simulation. Moreover, we consider both signs of $\omega$ and $m$. This provides a better understanding of the structure of the spectrum. As we see, the QBSs branch off from the QNM line associated with the outer light ring (see Fig.~1). In addition, the symmetry $(\omega,m) \to (-\omega^*, -m)$ is made manifest. In Table~\ref{QNMtable}, we show the numerical values of modes observed in Fig.~2, in order to compare the three methods more precisely. 

\begin{table}[H]
\begin{scriptsize}
\begin{center}
\begin{tabular}{c|c|c|c}
$m$ & WKB & CFM & MOL \\
\hline
+2 & $0.102+0.0000252i$ & $0.111 - 0.00000125i$ & $0.111-0.000130i$ \\
\hline
-1 & $0.0639-0.0718i$ & $0.114 -0.0610i$ & $0.114-0.0613i$ \\
\hline
-3 & $0.128-0.000118i $ & $0.131 - 0.0000982i$ & $0.131-0.000255i$ \\
\hline
-4 & $0.309-0.00746i$ & $0.307  -0.00490i$ & $0.307-0.00555i$ \\
\hline
-5 & $0.425-0.0826i$ & $0.450  -0.0242i$ & $0.449-0.0255i$ \\
\hline
-6 & $0.297-0.00000206i$ & $0.298-0.00000181i$ & $0.298-0.00650i$ \\
\end{tabular}
\end{center}
\end{scriptsize}
\caption{Numerical values (to 3 s.f.) of the QNM/QBS frequencies obtained from the three different methods for $C/cr_0=3,D/cr_0=0.1$. We observe good quantitative agreement between the CFM and MOL simulations. Discrepancies for $m=+2,-3,-6$ are a limitation of our time domain method for small imaginary parts. For $m = +2$, the approximations used to derive Eq.~(13) in the text lead to an incorrect prediction for the sign of the imaginary part using the WKB method. However, both the CFM and MOL shows that this mode is stable, with negative imaginary part.}
\label{QNMtable}
\end{table}

\end{document}